\documentclass[APA,LATO1COL]{WileyNJD-v2}

\usepackage{amssymb}
\usepackage{graphicx}
\usepackage{enumerate}
\usepackage{amsthm}
\usepackage{url} 
\usepackage{mathtools}
\usepackage{bbm}
\usepackage{dsfont}
\usepackage{amsfonts}
\usepackage{bm}
\usepackage{tablefootnote}
\usepackage[thinc]{esdiff}
\usepackage{hyperref}
\hypersetup{colorlinks=true,linkcolor=blue, citecolor=blue, linktocpage}
\mathtoolsset{showonlyrefs}
\allowdisplaybreaks

\newcommand{\bmu}{\bm{\mu}}
\newcommand{\bgamma}{\bm{\gamma}}
\newcommand{\bbeta}{\bm{\beta}}
\newcommand{\bmeta}{\bm{\eta}}
\newcommand{\bz}{\bm{z}}
\newcommand{\bw}{\bm{w}}
\newcommand{\p}{\textup{P}}
\newcommand{\e}{\textup{E}}
\newcommand{\bSigma}{\bm{\Sigma}}
\newcommand{\blambda}{\bm{\lambda}}
\newcommand{\bv}{\bm{v}}
\newcommand{\bxi}{\bm{\xi}}
\newcommand{\bphi}{\bm{\phi}}
\newcommand{\bmm}{\bm{m}}
\newcommand{\bzeta}{\bm{\zeta}}
\newcommand{\diag}{\textup{diag}}

\articletype{}%

\begin{document}

\title{A Multilayer Correlated Topic Model}

\author[1]{Ye Tian}

%

\authormark{Ye Tian}

\address[1]{\orgdiv{Department of Statistics}, \orgname{Columbia University}, \orgaddress{\state{New York}, \country{USA}}}
%
%

\corres{Ye Tian, \orgdiv{Department of Statistics}, \orgname{Columbia University}, \orgaddress{\state{New York}, \country{USA}}. \email{ye.t@columbia.edu}}
%

\abstract[Summary]{We proposed a novel multilayer correlated topic model (MCTM) to analyze how the main ideas inherit and vary between a document and its different segments, which helps understand an article's structure. The variational expectation-maximization (EM) algorithm was derived to estimate the posterior and parameters in MCTM. We introduced two potential applications of MCTM, including the paragraph-level document analysis and market basket data analysis. The effectiveness of MCTM in understanding the document structure has been verified by the great predictive performance on held-out documents and intuitive visualization. We also showed that MCTM could successfully capture customers' popular shopping patterns in the market basket analysis.}

%
%
\maketitle
%
%

\section{Introduction}\label{sec: intro}
Probabilistic topic modeling has been a popular research topic due to its effectiveness in learning the documents' underlying semantic structure. Since the latent Dirichlet allocation (LDA) came out \citep{blei2003latent}, researchers have developed a number of topic modeling methods, including the author-topic model \citep{rosen2012author}, hierarchical LDA model \citep{griffiths2004hierarchical}, the correlated topic model (CTM) \citep{blei2007correlated} and the dynamic topic model \citep{blei2006dynamic}. One thing that is ignored by many topic models but very important is that documents come with structures: a document can have paragraphs that contain sentences. Modeling such a local structure of an article can be very helpful to understand the document. An article can carry out some main ideas, while each paragraph is organized to introduce only a part of these ideas. Ideas of paragraphs vary around the main idea of the whole document \citep{du2010segmented}. Using the language of topic modeling, segments of an article can have different topic proportions while sharing some similarities with the document-level topic proportion. A relative assumption of LDA and some other document-level topic models is the out-of-bag assumption, which imposes the exchangeability of words within each document. It is almost impossible for the long text to have this assumption because different segments can express slightly different ideas, although they belong to the same article. In summary, capturing heterogeneity within different article segments is quite useful for understanding the document structure.

There has been some previous work tackling this problem. We can divide these methodologies into two categories. Methods in the first category introduce two types of topics, which they call document-level topics (or super-topics) and word-level topics, respectively, and segments with different super-topics can enjoy different word-level distributions. The latent Dirichlet co-clustering (LDCC) proposed by \cite{shafiei2006latent} is a direct application of this idea. \cite{li2006pachinko} developed pachinko allocation model (PAM), using a directed acyclic graph to capture the hierarchical structure of topics in different layers. \cite{hou2017multi} applied a similar idea in a multilayer multi-view topic model (mlmv\_LDA) for video classification. They all belong to the first category. One issue of this family of methods is the difficulty in interpreting the super-topics, especially when there are multiple levels. Instead of assigning a super-topic to segments, approaches belonging to the second category connect different layers of an article by parameter passing. Within each layer of an article, the passed parameters enjoy some variety while are still relative to the same document-level parameter. Compared with the first category, this way turns out to be more natural, and the results are more comfortable to illustrate because there is only one type of topic. Example methods in this category include the segmented topic model based on Poisson-Dirichlet process \citep{du2010segmented}, the sequential LDA model \citep{du2010segmented} and the LDA model based on partition \citep{guo2019improved}. These methods are all designed based on the LDA model \citep{blei2003latent}.

In this work, we proposed a new multilayer correlated topic model (MCTM), which belongs to the aforementioned second category. It is based on CTM \citep{blei2007correlated} and the topic of each word is generated from a logistic normal distribution \citep{atchison1980logistic}. Motivated by the dynamic topic model \citep{blei2006dynamic}, MCTM varies the topic distribution in each layer of a document by passing the mean parameter of the normal distribution. The dynamic topic model passes the mean parameter along articles published at different times to capture the changing trend, which can be seen as a series of models. Segments belonging to the same node in the higher layer in MCTM share the same mean parameter, which is similar to a tree structure. This design also shares a similar idea as the famous random effect model \citep{laird1982random}, where observations of different people have the same fixed effect and different random effect. 

We highlight our contribution in two aspects. First, a simple multilayer correlated topic model (MCTM) was proposed to be applied in analyzing the document structure and how the main ideas of an article are organized in different segments. It relaxes the out-of-bag assumption by only imposing the exchangeability within each segment. A variational expectation-maximization (EM) algorithm was derived to estimate the posterior and parameters in MCTM. Second, we visualized the results by connecting the document and its paragraphs with related topics in high proportions, which intuitively show the similarity and heterogeneity with different paragraphs of the same article. Such visualization can be beneficial in the paragraph-level document analysis.

The remaining of this paper is organized as follows. Section \ref{sec: mctm} introduces the MCTM coupled with the document generation procedure and its graphical representation in detail. In Section \ref{sec: inf}, we present the variation EM algorithm for the parameter and posterior estimation in MCTM. Section \ref{sec: applications} briefly discusses two potential applications of MCTM, including the paragraph-level document analysis and the market basket analysis. We conduct the experiments and present the results of these two applications in Section \ref{sec: exp results}. Finally, we close this paper with the summary of our work and the future avenues in Section \ref{sec: discussion}. Details of the derivation for variational EM algorithm is summarized in Appendix \ref{sec: vem derivation}.

\section{A Multilayer Correlated Topic Model (MCTM)}\label{sec: mctm}
We first introduce some notations and terms used in this section. 
\begin{itemize}
	\item A corpus is a collection of $D$ documents, which are denoted by the words they contain as $\{\bm{w}_d\}_{d=1}^D$. The corpus is the top level of our hierarchical model.
	\item The $d$-th document consists of $S_d$ segments $\{\bm{w}_{ds}\}_{s=1}^{S_d}$ with no intersection. The document is the second level of the hierarchical model.
	\item The $s$-th segment of $d$-th document is a sequence of $N_{ds}$ words $\{w_{dsn}\}_{n=1}^{N_{ds}}$. The segment is the third level in the whole model.
	\item A word $w \in \{1,\ldots, W\}$ is the bottom level of the topic model. 
\end{itemize}

\begin{boxtext}
The generative procedure of our model can be described as follows (for document $d$).
\begin{enumerate}
	\item Choose the number of segments $S_d \sim \textup{Poisson}(\upsilon_S)$;
	\item Draw $\bgamma_d \sim N(\bmu, \bSigma)$;
	\item For each segment $s = 1:S_d$:
	\begin{enumerate}
		\item Draw $\bmeta_{ds} \sim N(\bgamma_d, \bSigma)$;
		\item Draw the number of words $N_{ds} \sim \textup{Poisson}(\upsilon_N)$;
		\item For each word $n = 1:N_{ds}$:
			\begin{enumerate}
				\item Choose a topic $z_{dsn} \sim \textup{Categorical}(f(\bmeta_{ds}))$, where $f$ is the logistic normal function;
				\item Choose a word $w_{dsn} \sim \textup{Categorical}(\bbeta_{z_{dsn}})$.
			\end{enumerate}
	\end{enumerate}
\end{enumerate}
\end{boxtext}
\indent

\begin{figure}[!h]
	\centering
	\includegraphics[width=0.9\textwidth]{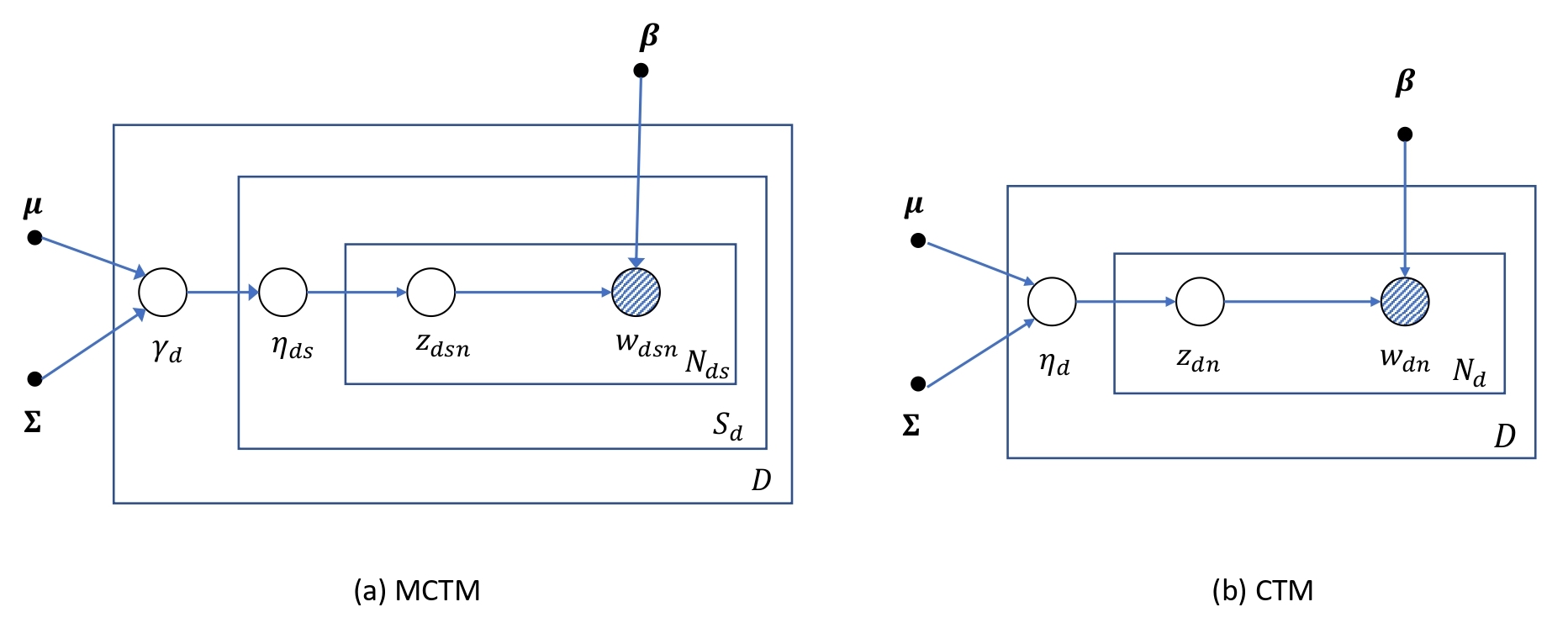}
	\caption{Graphical representations of MCTM (left) and CTM (right). The plates represent the replications of the same structure. The shaded nodes, unshaded nodes and black dots denote observed variables, hidden variables and hyperparameters, respectively.}
	\label{fig: mctm_ctm}
\end{figure}

The corresponding graphical representation of MCTM is shown in Figure \ref{fig: mctm_ctm}. The plates are used to represent the replications of the same structure. The shaded nodes, unshaded nodes and black dots denote observed variables, hidden variables and hyperparameters, respectively. Here we denote $\bbeta_{W \times K} = (\beta_{wz})_{W \times K}$, where $\beta_{wz} = \Pr(w_{dsn} = w|z_{dsn} = z)$ for any $d, s$ and $n$. $\bgamma$ and $\bz$ are hidden variables. $v_S$, $v_N$, $\bmu$, $\bSigma$ and $\bmeta$ are hyperparameters. In practice, we see paragraphs as different segments. As \cite{shafiei2006latent} pointed out, variables $\{N_{ds}\}_{d,s}$ are independent with all the other variables and we can observe $\{N_{ds}\}_{d,s}$ as well. Therefore without the loss of generality, the randomness $\{N_{ds}\}_{d,s}$ can be ignored. The topic $z_{dsn}$ corresponding to word $w_{dsn}$ given $\bmeta_{ds}$ is generated from a categorical distribution with parameter $f(\bmeta_{ds})$, where its $k$-th component
\begin{align}
	f^{(k)}(\bmeta_{ds}) = \frac{\exp(\eta_{dsk})}{\sum_{k=1}^K \exp(\eta_{dsk})}, k = 1, \ldots, K.
\end{align}
Given the hyperparameters $\bmu$, $\Sigma$ and $\bbeta$, the joint distribution of the document-level topic mixture parameters $\bgamma = \{\bgamma_d\}_{d=1}^D$, the segment-level topic mixture parameters $\{\bmeta_{ds}\}_{d,s}$, the word topics $\bm{z} = \{z_{dsn}\}_{d,s,n}$ and the observed words $\bm{w} = \{w_{dsn}\}_{d,s,n}$ is
\begin{align}
	p(\bgamma, \bmeta, \bz, \bw|\bmu, \bSigma, \bbeta) = \prod_{d=1}^Dp(\bgamma_d|\bmu, \bSigma)\prod_{s=1}^{S_d} p(\bmeta_{ds}|\bgamma_d, \bSigma) \prod_{n=1}^{N_{sd}} p(z_{dsn}|\bmeta_{ds})p(w_{dsn}|\bbeta, z_{dsn}).
\end{align}

The novel point of MCTM is that we consider one more layer than the CTM, which adds additional flexibility for the topic model, making it adapted to the nuanced analysis of a document in almost no more effort. More precisely, each segment has its parameter $\bmeta_{ds}$ to describe the topic distribution within that segment, which is expected to capture better the local topic distribution than the global parameter $\bgamma_d$. Simultaneously, MCTM does not require more hyperparameters than CTM because each segment uses that document's information to generate its parameter $\bmeta_{ds}$. As mentioned in the last section, MCTM enjoys analogous spirits with the famous random effect model \citep{laird1982random}. For each individual, the fixed population effect is the same, while the individual effect is random. Here we can see each document as a population and segments as individuals. 

It is essential to point out that MCTM can be of any depth if we pass the mean parameter of logistic normal distribution through each layer. It allows us to extend MCTM to more sophisticated structures such as the sentence-level topic model. In the following, we only consider MCTM exhibited in Figure \ref{fig: mctm_ctm} with one more layer than CTM and leave the arbitrarily deep models to future research.
 
\section{Inference}\label{sec: inf}
With the Bayesian model in hand, the next step is to develop the inference algorithm. The inference goal is to estimate the parameters in the model and compute the posterior, which is very useful in future prediction and model evaluation.

In many cases, especially for complicated models, it is impossible to carry out the exact inference. We have to rely on the Bayesian approximation approaches. There are three main families of approximation methods, including Markov Chain Monte Carlo (MCMC) \citep{robert2013monte}, variational inference (VI) \citep{blei2017variational} and expectation propagation \citep{minka2013expectation}. With MCMC and joint distribution, we can make the inference from complicated models. The calculation can be further simplified if the conjugacy holds. Among the MCMC family, one popular approach is Gibbs sampler \citep{gelfand1990sampling}. Gibbs sampler is available when the complete conditional distribution is known, which does not require the information about the joint distribution \citep{shafiei2006latent}. Besides, the conditional conjugacy is easier to be satisfied than the joint conjugacy \citep{blei2014build}.

VI is another family of Bayesian approximation methods \citep{jordan1999introduction, wainwright2008graphical}, which is an efficient inference method. It optimizes Kullback-Leibler (KL) divergence between exact posterior and the approximate one \citep{blei2017variational}. The simplest approximation family is the mean-field family, where it is assumed that all hidden variables are independent conditioned on observed data. The parameters used to describe the independent conditional distribution of hidden variables are called variational parameters. When the conditional conjugate priors are available, the coordinate ascent VI can be straightforward to conduct. \cite{blei2017variational}, and \cite{hoffman2013stochastic} discussed the relationship between the stepwise updates and the conditional posterior under the exponential family. There has been a myriad of variants of VI appearing in recent years, including the stochastic VI \citep{hoffman2013stochastic}, the automatic differentiation VI \citep{kucukelbir2015automatic, kucukelbir2017automatic}, and black box VI \citep{ranganath2014black}. Variational Expectation-Maximization (EM) algorithm also belongs to this family. It estimates the variational parameters and hyperparameters in two stages.

In this work, we follow similar steps of \cite{blei2007correlated} to develop the variational EM algorithm and update the estimates through the coordinate ascent algorithm. Denote the number of topics $K$ and documents $\{\bw_d\}_{d=1}^D$. Suppose the $d$-th document has $S_d$ segments. We are going to estimate the posterior $p(\bgamma, \bmeta, \bz|\bmu, \bSigma, \bbeta, \bw)$. Consider the mean-field variational family, where we approximate the posterior with
\begin{align}
	q(\bgamma, \bmeta, \bz|\blambda, \bv^2, \bxi, \bmm^2, \bphi) = \prod_{d=1}^D\prod_{k=1}^K q(\gamma_{dk}|\lambda_{dk}, v_{dk}^2) \prod_{d=1}^D\prod_{s=1}^{S_d}\prod_{k=1}^K q(\eta_{dsk}|\xi_{dsk}, m_{dsk}^2) \cdot \prod_{d=1}^D\prod_{s=1}^{S_d}\prod_{n=1}^{N_{ds}} q(z_{dsn}|\phi_{dsn}), \label{eq: q}
\end{align}
where $\gamma_{dk} \sim N(\lambda_{dk}, v_{dk}^2)$, $\eta_{dsk} \sim N(\xi_{dsk}, m_{dsk}^2)$ and $z_{dsn} \sim \textup{Multinomial}(\phi_{dsn})$. And $\bgamma, \bmeta, \bz$ are independent. The joint distribution of hidden variables
\begin{align}
	p(\bgamma, \bmeta, \bz, \bw|\bmu, \bSigma, \bbeta) &= \prod_{d=1}^D p(\bgamma_d|\bmu, \bSigma)\prod_{s=1}^{S_d} p(\bmeta_{ds}|\bgamma_d, \bSigma) \prod_{n=1}^{N_{sd}} p(z_{dsn}|\bmeta_{ds})p(w_{dsn}|\bbeta, z_{dsn}) \\
	&= \prod_{d=1}^D\frac{1}{(2\pi)^{\frac{K}{2}}|\bSigma|^{\frac{1}{2}}}\exp\left\{-\frac{1}{2}(\bgamma_d - \bmu)^T\bSigma^{-1}(\bgamma_d - \bmu)\right\} \prod_{s=1}^{S_d}\frac{1}{(2\pi)^{\frac{K}{2}}|\bSigma|^{\frac{1}{2}}}\exp\left\{-\frac{1}{2}(\bmeta_{ds} - \bgamma_d)^T\bSigma^{-1}(\bmeta_{ds} - \bgamma_d)\right\} \times \\
	&\quad \prod_{n=1}^{N_{sd}}\prod_{k=1}^K \left[f^{(k)}(\bmeta_{ds})\right]^{z_{dsn}^{(k)}}\prod_{w=1}^W \beta_{wk}^{z_{dsn}^{(k)}}.\label{eq: p}
\end{align}
In the E-step, we are going to find the estimate of $(\blambda,  \bm{v}^2, \bxi, \bmm^2, \bphi)$ to maximize the ELBO
\begin{align}
	\e_q[\log p(\bgamma, \bmeta, \bz, \bw|\bmu, \bSigma, \bbeta)] - \e_q[q(\bgamma, \bmeta, \bz|\blambda, \bm{v}^2, \bxi, \bmm^2, \bphi)],
\end{align}
which is shown to be a lower bound of the evidence $\sum_{d=1}^D\log p(\bw_{d}|\bmu, \bm{\Sigma}, \bbeta)$ \citep{blei2017variational}. The exact ELBO is still challenging to calculate, therefore we are going to find a lower bound of it and optimize the lower bound instead \citep{blei2007correlated}. To do this, we need to define additional variational parameters $\bzeta = \{\zeta_{ds}\}_{d,s=1}$. With some straightforward calculations, it can be shown that 
\begin{align}
	B(\bmu, \bSigma, \bbeta, \blambda, \bm{v}^2, \bxi, \bmm^2, \bphi) &= -\frac{D}{2}\log |\bm{\Sigma}| - \frac{KD}{2}\log 2\pi - \frac{1}{2}\textup{Tr}(diag(\bv_d^2)\bSigma^{-1})- \frac{1}{2}\sum_{d=1}^D(\blambda_d - \bmu)^T\bSigma^{-1}(\blambda_d - \bmu) - \frac{1}{2}(K\log 2\pi + \log |\bSigma| )\cdot \sum_{d=1}^D S_d \\
	&\quad- \frac{1}{2}\sum_{d=1}^D\sum_{s=1}^{S_d}\textup{Tr}(diag(\bv_d^2 + m_{ds}^2)\bSigma^{-1}) - \frac{1}{2}\sum_{d=1}^D\sum_{s=1}^{S_d}(\bxi_{ds} - \blambda_d)^T\bSigma^{-1}(\bxi_{ds} - \blambda_d) \\
	&\quad + \sum_{d=1}^D\sum_{s=1}^{S_d}\sum_{n=1}^{N_{ds}}\left[\sum_{k=1}^K \xi_{dsk}\phi_{dsnk} - \zeta_{ds}^{-1}\sum_{k=1}^K\exp(\xi_{dsk} + \frac{1}{2}m^2_{dsk}) + 1 - \log \zeta_{ds}\right] + \sum_{d=1}^D\sum_{s=1}^{S_d}\sum_{n=1}^{N_{ds}}\sum_{k=1}^K \phi_{dsnk}\log \beta_{w_{dsn}, k} \\
	&\quad + \frac{1}{2}\sum_{d=1}^D\sum_{k=1}^K(\log v_{dk}^2 + \log 2\pi + 1) + \frac{1}{2}\sum_{d=1}^D\sum_{s=1}^{S_d}\sum_{n=1}^{N_{ds}}\sum_{k=1}^K (\log m_{dks}^2 + \log 2\pi + 1) - \sum_{d=1}^D\sum_{s=1}^{S_d}\sum_{n=1}^{N_{ds}}\sum_{k=1}^K \phi_{dsnk} \\
	&\quad - \sum_{d=1}^D\sum_{s=1}^{S_d}\sum_{n=1}^{N_{ds}}\sum_{k=1}^K \phi_{dsnk}\log \phi_{dsnk}\\
	&\leq \e_q[\log p(\bgamma, \bmeta, \bz, \bw|\bmu, \bSigma, \bbeta)] - \e_q[q(\bgamma, \bmeta, \bz|\blambda, \bm{v}^2, \bxi, \bmm^2, \bphi)] \\
	&\leq \sum_{d=1}^D\log p(\bw_{d}|\bmu, \bm{\Sigma}, \bbeta).
\end{align}

The coordinate ascent algorithm can be easily conducted through stepwise optimization of $B(\bmu, \bSigma, \bbeta, \blambda, \bm{v}^2, \bxi, \bmm^2, \bphi)$. The updating formulas of estimate $(\hat{\zeta}_{ds}, \hat{\phi}_{dsnk}, \hat{\blambda}_d, \hat{v}_{dk})$ are
\begin{align}
	\hat{\zeta}_{ds} &= \sum_{k=1}^K \exp\left\{\hat{\xi}_{dsk} + \frac{1}{2}\hat{m}_{dsk}^2\right\}, \\
	\hat{\phi}_{dsnk} &\propto \beta_{w_{dsn}, k}\exp\{\hat{\xi}_{dsk}\}, k = 1, \ldots, K, \\
	\hat{\blambda}_d &= (S_d + 1)^{-1}\left(\sum_{s=1}^{S_d}\hat{\bxi}_{ds} + \bmu \right), \\
	\hat{v}_{dk} &= ((S_d + 1)(\bSigma^{-1})_{kk})^{-1}.
\end{align}
For $\bxi$, $\bmm$, we find their estimates through Newton-Raphson algorithm. The corresponding gradient and hessian matrix are
\begin{align}
	\nabla_{\bxi_{ds}}B &= -\bSigma^{-1}(\bxi_{ds} - \blambda_d) + \sum_{n=1}^{N_{ds}}\bphi_{dsn} -\frac{N_{ds}}{\zeta_{ds}}\exp\left\{\bxi_{ds} + \frac{1}{2}\bmm_{ds}^2\right\}, \\
	\nabla_{\bxi_{ds}}^2 B &= -\bSigma^{-1}  -\frac{N_{ds}}{\zeta_{ds}}\textup{diag}\left(\exp\left\{\bxi_{ds} + \frac{1}{2}\bmm_{ds}^2\right\}\right), \\
	\frac{\partial B}{\partial m_{dsk}^2} &=  -\frac{1}{2}(\Sigma^{-1})_{kk} - \frac{N_{ds}}{2\zeta_{ds}}\exp\left\{\bxi_{ds} + \frac{1}{2}\bmm_{ds}^2\right\} + \frac{1}{2m_{dsk}^2},\\
	\frac{\partial^2 B}{\partial (m_{dsk}^2)^2} &=   -\frac{N_{ds}}{4\zeta_{ds}}\exp\left\{\bxi_{ds} + \frac{1}{2}\bmm_{ds}^2\right\} - \frac{1}{2m_{dsk}^4}.
\end{align}
In the M-step, we update the hyperparameters $(\bmu, \Sigma, \bbeta)$ to maximize $B(\bmu, \bSigma, \bbeta, \blambda, \bm{v}^2, \bxi, \bmm^2, \bphi)$ each step by
\begin{align}
	\hat{\beta}_{wk} &\propto \sum_{d=1}^D \sum_{s=1}^{S_d} \#\{n: w_{sdn} = w\}\hat{\phi}_{dsnk},\\
	\hat{\bmu} &= \frac{1}{D}\sum_{d=1}^D \hat{\blambda}_d,
\end{align}
\begin{align}
	\hat{\bSigma} &= \frac{1}{D + \sum_{d=1}^D S_d}\cdot \left[\sum_{d=1}^D \diag (\hat{\bv}_d^2) + \sum_{d=1}^D\sum_{s=1}^{S_d}\diag (\hat{\bv}_d^2 + \hat{\bmm}_{ds}^2) + \sum_{d=1}^D(\hat{\blambda}_d - \hat{\bmu})(\hat{\blambda}_d - \hat{\bmu})^T + \sum_{d=1}^D\sum_{s=1}^{S_d}(\hat{\bxi}_{ds} - \hat{\blambda}_d)(\hat{\bxi}_{ds} - \hat{\blambda}_d)^T \right].  
\end{align}
The detailed derivation of the results above can be found in Appendix \ref{sec: vem derivation}.

\section{Applications}\label{sec: applications}
In this section, we point out two potential applications, including paragraph-level document analysis and market basket analysis, where MCTM can be directly applied. The experiment results on these two problems will be presented in the next section.

\subsection{Paragraph-level document analysis}
As mentioned in Section \ref{sec: intro}, the original LDA model \citep{blei2003latent} and CTM \citep{blei2007correlated} conducted only the document-level topic analysis, which assume the interchangeability of words within each document. MCTM relaxes this assumption by only requiring the interchangeability within each segment. In practice, each paragraph is natural to be a segment. An article has some main ideas, and each paragraph of it is organized and can carry out slightly different information around these ideas \citep{du2010segmented}. Understanding what each paragraph of an article is talking about can be very useful in text mining.

\subsection{Market basket analysis}\label{subset: MBA}
The decision process that consumers select items from various products in the same shopping trip is called a market basket choice problem \citep{russell1999multiple, russell2000analysis}. Numerous models have been developed to analyze market basket data. \cite{hruschka2014linking} fitted LDA and CTM on market basket data and obtained some interesting conclusions. Although the topic models might not be favorable in dealing with the market basket data \citep{hruschka2019comparing}, it is helpful for us to explore the common shopping pattern and the preference of each customer via topic models. 	The topic models' results could help the market design better promotion strategies and recommend appropriate products to customers.

We can adapt MCTM into the market basket analysis with no extra efforts. Items can be seen as different words in document analysis. Each topic indicates a specific shopping pattern of these items. Consider different customers $d = 1, \ldots D$ and the $d$-th customer has $s = 1, \ldots, S_d$ shopping trips in total. Within the $s$-th trip of customer $d$, a list of items $w_{ds1}, \ldots, w_{dsN_{ds}}$ are added into the cart. Each customer can be seen as a ``document,'' and each trip can be seen as a ``segment''. The data generation procedure is the same as presented in Section \ref{sec: mctm}. The corresponding variational EM algorithm presented in Section \ref{sec: inf} can be immediately applied.

\section{Experimental Results}\label{sec: exp results}
\subsection{NIPS dataset}
We first demonstrate the effectiveness of MCTM to capture the local structure of articles in paragraph-level document analysis. We take NIPS conference papers published during 2015-2017 as the dataset. We removed all the words appearing for less than three times and all the paragraphs with less than 20 words. Also, we used a list of scientific stop words \footnote{The list can be found at \url{https://github.com/seinecle/Stopwords/blob/master/src/scientific/en/scientificstopwords.txt}.} and common stop words from ISO dataset provided by the R package \texttt{stopwords}. Finally, the NIPS dataset contains 1646 documents with 1181156 words. There are 20667 unique words in total.

We evaluate different models by investigating the perplexity. Define the training set as $\bw_{\textup{train}}$ and the parameters fitted using the training set are $\Upsilon$. Denote the held-out documents as $\bw_{\textup{held-out}} = \{\bw_{d}\}_{d=1}^D$ and suppose $\bw_{d}$ has $N_d$ words in total. Then we define the perplexity as 
\begin{align}
	\textup{Perplexity} = \left(\prod_{d=1}^D \p(\bw_{d}|\Upsilon, \bw_{\textup{train}})\right)^{-1/(\sum_{d=1}^D N_d)}.
\end{align}
The lower perplexity denotes a more powerful model. In experiments, it is estimated by following the harmonic mean method and treating the approximate posteriors as the real proposed ones \citep{wallach2009evaluation}.

We used the 2016-2017 NIPS papers, which contains 1244 documents and 930588 words in total, as the training set. The 2015 NIPS papers, which contains 402 documents and 250568 words, are used as held-out documents. We fitted the LDA model \citep{blei2003latent}, CTM \citep{blei2007correlated} and MCTM by considering 10, 20, 30, 40, 50, 60, 70, 80 topics. The LDA model and CTM are implemented by R package \texttt{topicmodels} \citep{hornik2011topicmodels}. The inference algorithm is run until the relative change of ELBO or the lower bound of ELBO is less than 0.0001\%. 

\begin{table}[!htp]
\resizebox{\columnwidth}{!}{%
  \begin{tabular}{ccccccccc}
    topic 1 & topic 2 &topic 3 &topic 4 &topic 5  & topic 6 &topic 7 &topic 8 &topic 9\\
    ``optimization" &``classification" &``memory" &``image processing" &``graph" & ``bound" &``kernel" &``online learning" &``acknowledgment"\\
    \hline
gradient  & loss  & size  & images  & graph  & bound  & kernel  & regret  & supported \\ 
vector  & classifier  & memory  & image  & nodes  & bounds  & kernels  & algorithms  & grant \\ 
optimization  & classifiers  & compression  & layers  & node  & upper  & space  & online  & nsf \\ 
loss  & examples  & bit  & dataset  & edges  & bounded  & gaussian  & bandit  & acknowledgments \\ 
descent  & binary  & quantization  & neural  & graphs  & inequality  & rkhs  & action  & reward \\ 
step  & risk  & values  & networks  & vertices  & holds  & hilbert  & reward  & agent \\ 
iteration  & hinge  & cost  & convolutional  & clustering  & tight  & functions  & learner  & university \\ 
functions  & losses  & datasets  & architecture  & tree  & terms  & reproducing  & actions  & tasks \\ 
respect  & decision  & dataset  & features  & algorithms  & implies  & matrix  & settings  & policies \\ 
approximation  & neural  & hash  & size  & sets  & constants  & features  & unknown  & neural \\      
\hline
  \end{tabular}}
  \caption{Top 10 words with highest frequency from 9 selected topics. The results is obtained from a 40-topic MCTM.}
  \label{table: topic words}
\end{table}

\begin{figure}[!htp]
\centering
  \includegraphics[width=0.9\textwidth]{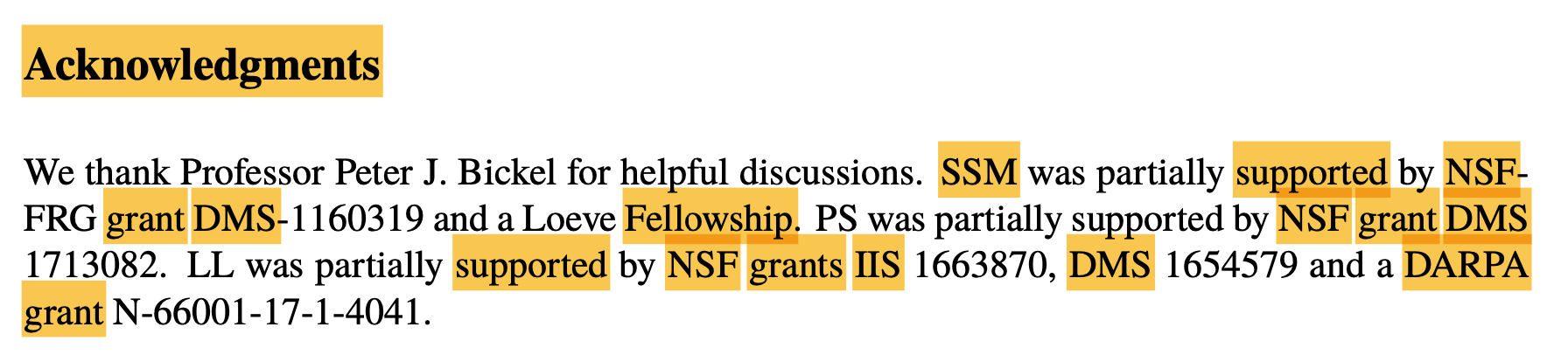}
  \caption{The last paragraph of 2017 NIPS paper ``On clustering network-valued data". Training words are highlighted. In this paragraph the ``acknowledgement" topic proportion is 96.6\% while in the whole paper it is only 0.0027\%. }
  \label{figure: paper 2017 example}
\end{figure}

First, we presented 9 topics with the corresponding top 10 words with the highest frequency derived from the training data by fitting a 40-topic MCTM in Table \ref{table: topic words}. Notice that MCTM indeed captures the hidden structure of the documents, and these topics are interpretable. Notice that there is a special topic, which we call ``acknowledgment", including some words often used in the acknowledgment paragraph. Our model can capture this topic because of its hierarchical structure and flexible variation between documents and paragraphs. To intuitively show how MCTM captures the difference between topics of different paragraphs, we take the following example. Figure \ref{figure: paper 2017 example} shows the last paragraph of 2017 NIPS paper ``On clustering network-valued data" \citep{mukherjee2017clustering}. This paper is one of the training documents, and we highlight the training words in this paragraph. This paragraph's approximate posterior topic distribution shows that in 96.6\% probability, a word is generated from the ``acknowledgment" topic. While in the whole paper, this probability is only 0.0027\%. Capturing the difference between the document-level topic proportion and paragraph-level proportion leads to the success of MCTM when we are interested in the local structure of an article.

\begin{figure}[!htp]
\centering
  \includegraphics[width=0.6\textwidth]{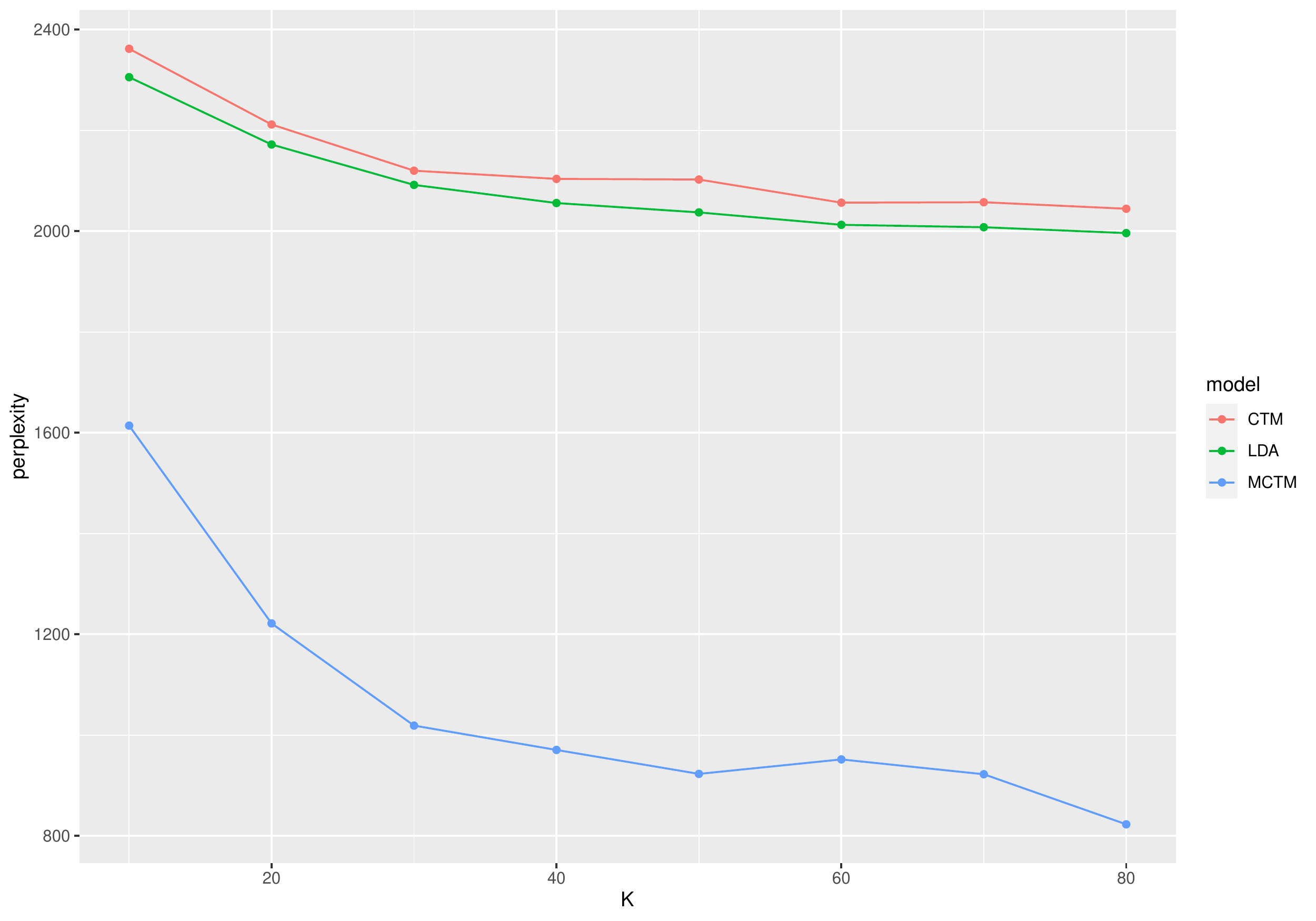}
  \caption{Perplexity of LDA, CTM and MCTM with respect to different number of topics and percentage of observed words in held-out documents.}
  \label{figure: perp_held_out}
\end{figure}

Next, we presented the perplexity with respect to different numbers of topics $K$ in Figure \ref{figure: perp_held_out}. From the results it can be observed that LDA and CTM achieve comparable performance while MCTM outperforms both of them. This is because MCTM can capture more sophisticated document structure which LDA and CTM cannot describe.

\begin{figure}[!h]
\centering
  \includegraphics[width=1.03\textwidth]{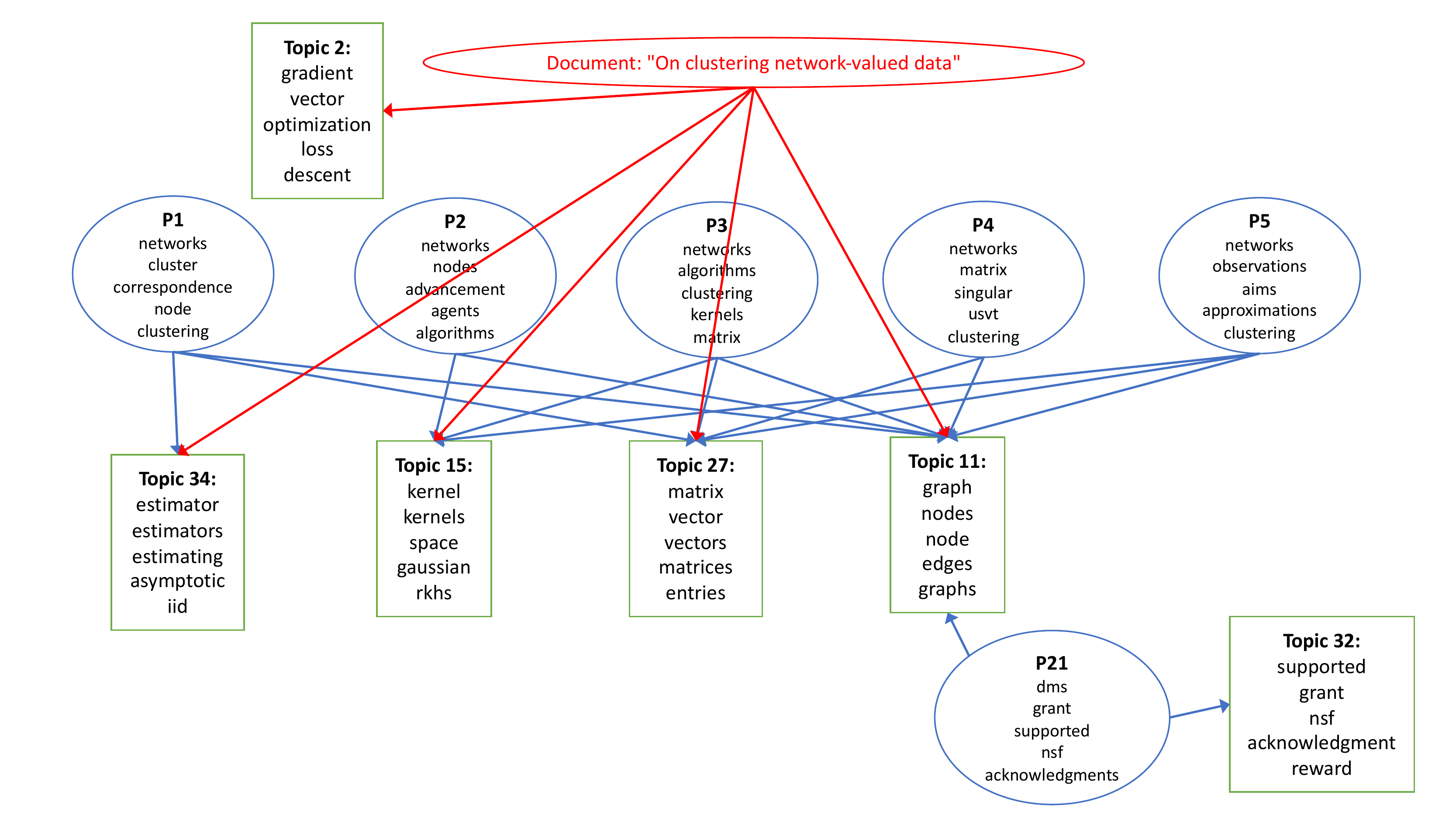}
  \caption{Document ``On clustering network-valued data" with its several paragraphs and some topics. The document and its paragraphs are connected with topics achieving over $1\%$ proportion level.}
  \label{figure: docu3}
\end{figure}

Finally, we visualize the results of MCTM by constructing an intuitive figure of the document ``On clustering network-valued data" with its several paragraphs and topics. It is exhibited as Figure \ref{figure: docu3}. Each circle represents a paragraph of the document with 5 highest-frequency words in that paragraph, while each rectangle denotes a topic with 5 top words. We connect the document and its paragraphs with topics achieving over $1\%$ proportion level. It can be observed that the document has connections with topics 2, 11, 15, 27, and 34, while some paragraphs only focus on some of them. For instance, paragraph 4 discusses mainly the ``graph'' and ``matrix''. All six paragraphs are connected with topic 11 ``graph'', which is the whole document's focus. Besides, paragraph 21 only connects with topic 11 ``graph'' and topic 32 ``acknowledgment''. We have shown this paragraph in Figure \ref{figure: paper 2017 example}, and it is an acknowledgment paragraph. Such a figure can be immediately drawn for each document after fitting MCTM, which clearly shows the local structure.

\subsection{Instacart market basket dataset}
In this section, we apply MCTM to study the market basket data by following the idea described in Section \ref{subset: MBA}. The dataset we used is the shopping data provided by Instacart, a big grocery market in the United States and Canada. It consists of the records of over 200000 customers. We take the first 1000 customers, coupled with 15145 trips and 150150 bought items as our training data. On each trip, the customer bought at least four items. There are 134 different categories to which the products belong. We fit MCTM based on these categories instead of single items.

\begin{table}[!htp]
\centering
\resizebox{\columnwidth}{!}{%
  \begin{tabular}{ccccc}
    topic 1 & topic 2 &topic 3 &topic 4 &topic 5\\
    ``laundry and other household" &``baby and body care" &``frozen products" &``vegetable, fruits and milk" &"drinking and body care"\\
    \hline
laundry  & baby food formula  & frozen meals  & fresh vegetables  & body lotions soap \\ 
deodorants  & diapers wipes  & frozen breakfast  & fresh fruits  & beers coolers \\ 
mint gum  & baby bath body care  & frozen pizza  & packaged vegetables fruits  & muscles joints pain relief \\ 
cleaning products  & energy granola bars  & instant foods  & yogurt  & dog food care \\ 
dish detergents  & fruit vegetable snacks  & cereal  & packaged cheese  & white wines \\ 
trash bags liners  & laundry  & frozen appetizers sides  & chips pretzels  & air fresheners candles \\ 
oral hygiene  & canned fruit applesauce  & plates bowls cups flatware  & water seltzer sparkling water  & ice cream toppings \\ 
baby accessories  & food storage  & frozen dessert  & frozen produce  & red wines \\ 
more household  & cereal  & frozen vegan vegetarian  & ice cream ice  & first aid \\ 
paper goods  & hair care  & hair care  & milk  & spirits \\ \hline
  \end{tabular}}
  \caption{5 selected topics and the corresponding top 10 categories from derived from a 20-topic MCTM.}
  \label{table: topic shopping}
\end{table}

We fitted a 20-topic MCTM and presented 5 topics and corresponding top 10 categories in Table \ref{table: topic shopping}. Observe that the MCTM successfully captures some shopping needs, including the household, baby care, body care, frozen products, vegetables, and drinking. Besides, by looking at each customer's topic proportion, we can better know the categories each customer often buys, which helps analyze each consumer's preference and promote the right products to them. The analysis of individual customers' behavior is out of our discussion scope, and we do not cover it here. 

\section{Discussion}\label{sec: discussion}
In this work, we proposed an innovative multilayer correlated topic model (MCTM) to analyze the documents' local structure. It is motivated by the connection between articles published at different times in the dynamic topic model \citep{blei2006dynamic} and the group structure used in the random effect model \citep{laird1982random}. The variational EM algorithm, coupled with MCTM, was derived for posterior and parameter estimation. We introduced two potential applications of MCTM, including the paragraph-level document analysis and market basket data analysis. In numerical studies, we demonstrated the effectiveness of MCTM in capturing the document structure, and MCTM was confirmed to improve LDA and CTM. We visualized the relationship between the main ideas of a document and its paragraphs. Besides, we captured some hidden shopping patterns among customers by applying MCTM on market basket data. 

Our work's main contribution is that we developed a very natural and straightforward way to extend CTM \citep{blei2007correlated} in almost no extra effort, making it possible to capture the heterogeneity between paragraphs of a document. Understanding the local structure of an article can help investigate how central ideas are organized inside a document. 

There are a few future avenues that are interesting to explore. The first one is the usage of the covariance matrix in the logistic normal distribution. As \cite{blei2007correlated} suggested, the covariance can help analyze the pairwise correlation between different topics. Similar results can be obtained for MCTM as well. Besides, in our framework, topics do not have to share the same correlation structure at different levels. Using separate covariance matrices at different levels might help interpret the relationship between a document and its paragraphs. Second, as we mentioned before, MCTM can be of any depth, making it ready to be extended for more sophisticated modeling such as the sentence-level topic modeling. Another appealing direction is to look for better ways to visualize the results of MCTM. Figure \ref{figure: docu3} provides a good example, and we expect to see more intuitive visualization methods to display the results.

\section*{Data accessibility}
The NIPS dataset is available at \url{https://www.kaggle.com/benhamner/nips-papers}, and the Instacart market basket dataset is available at \url{https://www.kaggle.com/c/instacart-market-basket-analysis}.

\section*{Acknowledgments}
The author's research is sponsored by Dean's Fellowship of Graduate School of Arts and Sciences, Columbia University.

\appendix

\section{Derivation of variational EM algorithm in Section 3}\label{sec: vem derivation}
In this section, we treat $z_{dsn}$ as a $K$-vector as $z_{dsn} = (z_{dsn}^{(1)}, \ldots, z_{dsn}^{(K)})^T$, where each coordinate is an indicator. For instance, if the $n$-th word in the $s$-th segment of the $d$-th document has topic $k$, then $z_{dsn}^{(j)} = \mathds{1}(j =k)$ for $j = 1, \ldots, K$. 

By \cite{blei2017variational}, the ELBO equals to
\begin{align}
	\e_q[\log p(\bgamma, \bmeta, \bz, \bw|\bmu, \bSigma, \bbeta)] - \e_q[q(\bgamma, \bmeta, \bz|\blambda, \bm{v}^2, \bxi, \bmm^2, \bphi)].
\end{align}
Due to \eqref{eq: p} and the normal density, we have
\begin{align}
	\e_q[\log p(\bgamma_d|\bmu, \bSigma)] &= \frac{1}{2}\log |\bSigma|^{-1} - \frac{K}{2}\log 2\pi - \frac{1}{2}\e_q[(\bgamma_d -\bmu)^T\bSigma^{-1}(\bgamma_d -\bmu)] \\
	&= -\frac{1}{2}\log |\bSigma| - \frac{K}{2}\log 2\pi - \frac{1}{2}\textup{Tr}(\textup{diag}(\bv_d^2)\bSigma^{-1}) - \frac{1}{2} (\blambda_d -\bmu)^T\bSigma^{-1}(\blambda_d -\bmu).\label{eq 1}
\end{align}
Similarly, it holds that
\begin{align}
	\e_q[\log p(\bmeta_{ds}|\bgamma_d, \bSigma)] &= -\frac{1}{2}\log |\bSigma| - \frac{K}{2}\log 2\pi- \frac{1}{2}\e_q[(\bmeta_{ds} -\bgamma_d)^T\bSigma^{-1}(\bmeta_{ds} -\bgamma_d)] \\
	&= -\frac{1}{2}\log |\bSigma| - \frac{K}{2}\log 2\pi - \frac{1}{2}\textup{Tr}(\textup{diag}(\bmm_{ds}^2 + \bv_d^2)\bSigma^{-1}) - \frac{1}{2} (\bxi_{ds} -\blambda_d)^T\bSigma^{-1}(\bxi_{ds} -\blambda_d). \label{eq 2}
\end{align}
According to the categorical distribution, we have
\begin{align}
	\e_q[\log p(z_{dsn}|\bmeta_{ds})] = \e_q(\bmeta_{ds}^T \bz_{dsn}) - \e_q\left[\log \left(\sum_{k=1}^K \exp (\eta_{dsk})\right)\right].
\end{align}
Since the second term does not have a close form, we follow the strategy of \cite{blei2007correlated} by introducing additional positive parameters $\bxi = \{\xi_{ds}\}_{d,s}$. It can be easily checked that
\begin{align}
	\e_q\left[\log \left(\sum_{k=1}^K \exp (\eta_{dsk})\right)\right] &\leq \xi_{ds}^{-1}\left[\sum_{k=1}^K\e_q(\exp (\eta_{dsk}))\right] -1 + \log \xi_{ds} \\
	&=  \xi_{ds}^{-1}\left[\sum_{k=1}^K\e_q\left(\exp \left(\xi_{dsk}+\frac{1}{2}m_{dsk}^2\right)\right)\right] -1 + \log \xi_{ds},\label{eq 3}
\end{align}
where the second equality comes from the moment generating function of normal distributed $\xi_{dsk}$ under $q$. In addition, we have
\begin{align}
	\e_q[\log p(w_{dsn}|z_{dsn}, \bbeta)] = \sum_{k=1}^K \phi_{dsnk}\log\beta_{w_{dsn},k}.  \label{eq 4}
\end{align}

On the other hand, it holds that
\begin{align}
	\e_q [\log q(\gamma_{dk}|\lambda_{dk}, v^2_{dk})] &= -\frac{1}{2}(\log v^2_{dk} + \log 2\pi + 1), \label{eq 5}\\
	\e_q [\log q(\eta_{dsk}|\xi_{dsk}, m^2_{dsk})] &= -\frac{1}{2}(\log m^2_{dsk} + \log 2\pi + 1),\label{eq 6}\\
	\e_q[\log q(z_{dsn}|\phi_{dsn})] &= \phi_{dsnk}\log \phi_{dsnk}.\label{eq 7}
\end{align}
Combining \eqref{eq: p}, \eqref{eq: q}, \eqref{eq 1}, \eqref{eq 2}, \eqref{eq 3}, \eqref{eq 4}, \eqref{eq 5}, \eqref{eq 6}, \eqref{eq 7}, it follows that ELBO can be bounded by $B(\bmu, \bSigma, \bbeta, \blambda, \bm{v}^2, \bxi, \bmm^2, \bphi)$. Then the derivations exhibited in Section \ref{sec: inf} are straightforward to obtain.

\bibliography{wileyNJD-APA}%

\nocite{*}

\end{document}